\documentclass{article}
\usepackage{amssymb}

\usepackage{graphicx}
\usepackage{amsmath}

\input{tcilatex}

\begin{document}

\begin{center}
\textbf{PHENOMENON OF THE TIME-REVERSAL-VIOLATING PHOTON POLARIZATION PLANE
ROTATION BY A GAS PLACED TO AN ELECTRIC FIELD}
\end{center}

\bigskip

\begin{center}
V.G.Baryshevsky
\end{center}

\bigskip

1. Violation of time reversal symmetry has been observed only in $K_{0}$%
-decay many years ago \cite{1}, and remains one of the great unsolved
problems in elementary particle physics. Since the discovery of the
CP-violation in decay of $K_{0}$-mesons, a few attempts have been undertaken
to observe this phenomenon experimentally in different processes. However,
that experiments have not been successful. At the present time novel more
precise experimental schemes are actively discussed: observation of the atom 
\cite{2} and neutron \cite{3} electric dipole moment, T-violating (time
reversal) atom (molecule) spin rotation in a laser wave and T-violating
optical activity of an atomic or molecular gas \cite{4,5}.

According to \cite{6} the violation of the time reversal symmetry stipulates
the appearance of the new optical phenomenon: the \ photon polarization
plane rotation and circular \ dichroism in an optical homogeneous isotropic
matter placed to an electric field. This T-odd phenomenon is the kinematic
analog of the famous T-even phenomenon of the photon polarization plane
rotation in a matter placed to a magnetic field (Faraday phenomenon).

In the present paper T-odd phenomenon of the photon polarization plane
rotation (circular dichroism) is considered for an atomic (molecular) gas
placed to an electric field.

The expression for the T non-invariant polarizability of an atom (molecule)
placed to an electric field is obtained. It is shown that the T-odd plane
rotation angle $\vartheta _{T}$ increases when the interaction energy of an
atom (molecule) with an electric field is the same order as the {\Large \ }%
opposite parity levels spacing.

2. In accordance with \cite{4,5,6}\ the T-reversal violating dielectric
permittivity tensor $\varepsilon _{ik}$\ for an optical diluted matter ( $%
\varepsilon _{ik}-\delta _{ik}\ll 1$, \ $\delta _{ik}$\ is the Kronecker
symbol) is given by

\begin{equation}
\varepsilon _{ik}=\delta _{ik}+\chi _{ik}=\delta _{ik}+\frac{4\pi \rho }{%
k^{2}}f_{ik}(0),
\end{equation}
where $\chi _{ik}$\ is the polarizability tensor of a matter, $\rho $\ is
the number of atoms (molecules) per $cm^{3}$, $k$\ is the photon wave
number; $f_{ik}(0)$\ is the tensor part of the zero angle amplitude of
elastic coherent scattering of a photon by an atom (molecule) $%
f(0)=f_{ik}(0)e_{i}^{\prime \ast }e_{k}$. Here $\overrightarrow{e}$\ and $%
\overrightarrow{e}^{\prime }$ \ are the polarization vectors of initial and
scattered photons. Indices $i=1,2,3$ are referred to coordinates $x,y,z$\
respectively, repeated indices imply summation.

Let photons are scattered by nonpolarized atoms (molecules) interacted with
an electric field $\overrightarrow{E}$.

The amplitude $f_{ik}(0)$\ can be written as \cite{6}:

\begin{equation}
f_{ik}(0)=f_{ik}^{ev}+\frac{\omega ^{2}}{c^{2}}[i\beta _{s}^{P}\varepsilon
_{ikl}n_{\gamma l}+i\beta _{E}^{T}\varepsilon _{ikl}n_{El}],
\end{equation}
where $f_{ik}^{ev}$\ is the T, P even (invariant) part of \ $f_{ik}(0)$\ , $%
\beta _{s}^{P}$\ is the scalar P-violating polarizability of an atom
(molecule), $\beta _{E}^{T}$\ is the scalar T and P violating polarizability
of an atom (molecule), $\varepsilon _{ikl}$\ is the totally antisymmetrical
unit tensor of the rank three, $\overrightarrow{n}_{\gamma }=\dfrac{%
\overrightarrow{k}}{k}$\ , $\overrightarrow{k}$\ is the photon wave vector, $%
\overrightarrow{n}_{E}=\dfrac{\overrightarrow{E}}{E}$\ .

The term proportional to $\beta _{s}^{P}$ describes the T-invariant
P-violating light polarization plane rotation (and circular dichroism) in a
gas \cite{7}.

The corresponding refractive index $N$\ can be written in the following form
convenient for the further analysis:

\begin{equation}
N=1+\frac{2\pi \rho }{k^{2}}\left\{ f_{ik}^{ev}(0)e_{i}^{\ast }e_{k}+i\frac{%
\omega ^{2}}{c^{2}}(\beta _{s}^{P}\overrightarrow{n}_{\gamma }+\beta _{E}^{T}%
\overrightarrow{n}_{E})[\overrightarrow{e}^{\ast }\times \overrightarrow{e}%
]\right\} .
\end{equation}

The unit vectors describing the circular polarization of photons are: $%
\overrightarrow{e}_{+}=-\dfrac{\overrightarrow{e}_{1}+i\overrightarrow{e}_{2}%
}{\sqrt{2}}\ $for the right and, $\overrightarrow{e}_{-}=\dfrac{%
\overrightarrow{e}_{1}-i\overrightarrow{e}_{2}}{\sqrt{2}}$\ for the left
circular polarization, where $\overrightarrow{e}_{1}\perp \overrightarrow{e}%
_{2}\ $, $\overrightarrow{e}_{2}=[\overrightarrow{n}_{\gamma }\times 
\overrightarrow{e}_{1}]\ $are the unit polarization vectors of a linearly
polarized photon, $[\overrightarrow{e}_{1}\times \overrightarrow{e}_{2}]=%
\overrightarrow{n}_{\gamma }$, $\overrightarrow{e}_{1}=-\dfrac{%
\overrightarrow{e}_{+}-\overrightarrow{e}_{-}}{\sqrt{2}}$, $\overrightarrow{e%
}_{2}=-\dfrac{\overrightarrow{e}_{+}+\overrightarrow{e}_{-}}{i\sqrt{2}}$.

Let an electromagnetic wave propagates in a gas along the direction of the
electric field\ $\overrightarrow{E}$.\ \ \ \ \ \ \ \ 

The refractive indices for the right $N_{+}$\ and the left $N_{-}$\ circular
polarized photons can be written as:

\begin{equation}
N_{\pm }=1+\frac{2\pi \rho }{k^{2}}f_{\pm }(0)=1+\frac{2\pi \rho }{k^{2}}%
\left\{ f^{ev}(0)\mp \frac{\omega ^{2}}{c^{2}}\left[ \beta _{s}^{P}+\beta
_{E}^{T}(\overrightarrow{n}_{E}\overrightarrow{n}_{\gamma })\right] \right\}
,
\end{equation}
where $f_{+}(0)(f_{-}(0))$ is the zero angle amplitude of the elastic
coherent scattering of a right (left) circular polarized photon by an atom
(molecule).

Let photons with the linear polarization $\overrightarrow{e}_{1}=-\dfrac{%
\overrightarrow{e}_{+}-\overrightarrow{e_{-}}}{\sqrt{2}}$\ fall in a gas.
The polarization vector of a photon in a gas $\overrightarrow{e}_{1}^{\prime
}$\ can be written as: 
\begin{eqnarray}
\overrightarrow{e}_{1}^{\prime }\ &=&-\frac{\overrightarrow{e}_{+}\ }{\sqrt{2%
}}e^{ikN_{+}L}+\frac{\overrightarrow{e}_{-}\ }{\sqrt{2}}e^{ikN_{-}L}= \\
&=&e^{\frac{1}{2}ik(N_{+}+N_{-})L}\left\{ \overrightarrow{e}_{1}\cos \frac{1%
}{2}k(N_{+}-N_{-})L-\overrightarrow{e}_{2}\sin \frac{1}{2}%
k(N_{+}-N_{-})L\right\} ,  \notag
\end{eqnarray}
where $L$\ is the photon propagation length in a medium.

As one can see, the photon polarization plane rotates in a gas. The angle of
rotation $\vartheta $\ is 
\begin{eqnarray}
\vartheta &=&\frac{1}{2}k\func{Re}(N_{+}-N_{-})L=\frac{\pi \rho }{k}\func{Re}%
\left[ f_{+}(0)-f_{-}(0)\right] = \\
&=&-\frac{2\pi \rho \omega }{c}\left[ \beta _{s}^{P}+\beta _{E}^{T}(%
\overrightarrow{n}_{E}\overrightarrow{n}_{\gamma })\right] ,  \notag
\end{eqnarray}
where $\func{Re}N_{\pm }$\ is the real part of $N_{\pm }$.

It should be noted that $\vartheta >0$\ corresponds to the right and $%
\vartheta <0$\ corresponds to the left rotation of the light polarization
plane, where the right (positive) rotation is the rotation recording by the
light observer as the clockwise one.

In accordance with (6) the T-odd interaction results in the photon
polarization plane rotation around the electric field $\overrightarrow{E}$%
{\LARGE \ }direction. The angle of rotation is proportional to the
polarizability $\beta _{E}^{T}$ and $\overrightarrow{E},\overrightarrow{n}%
_{\gamma }$ correlation. Together with the T-odd effect there is \ the
T-even P-odd polarization plane rotation phenomenon determining by the
polarizability $\beta _{s}^{P}$ and independent on the $\overrightarrow{E},%
\overrightarrow{n}_{\gamma }$ correlation. The T-odd rotation dependence on
the electric field {\LARGE \ }$\overrightarrow{E}$ orientation relatively to
the $\overrightarrow{n}_{\gamma }$ direction allows to distinguish T-odd and
T-even P-odd phenomena experimentally.

The refractive indices $N_{\pm }$\ have both real and imaginary parts. The
imaginary parts of the refractive indices $(\func{Im}N_{\pm }\sim \func{Im}%
\beta _{E}^{T}(\overrightarrow{n}_{E}\overrightarrow{n}_{\gamma }))$\ are
responsible for the T-reversal violating circular dichroism. Due to the this
process linearly polarized photons take circular polarization. The sign of
the circular polarization depends on the sign of scalar production $(%
\overrightarrow{n}_{E}\overrightarrow{n}_{\gamma })$\ that allows to
separate T-odd circular dichroism from P-odd T-even circular dichroism. The
last one is proportional to $\func{Im}\beta _{s}^{P}$.

3.{\Large \ }In order to estimate the magnitude of the effect one should
obtain the T-odd polarizability $\beta _{E}^{T}$\ or (that is actually the
same, see (2,6)) the amplitude $f_{\pm }(0)$ of elastic coherent scattering
of a photon by an atom (molecule).

According to quantum electrodynamics the elastic coherent scattering at the
zero angle can be considered as the succession of two processes: the first
one is the absorption of the initial photon (momentum $\overrightarrow{k}$)
and the transition of the atom (molecule) from the initial state $\left|
N_{0}\right\rangle $ with the energy $E_{N_{0}}$ to an intermediate state $%
\left| F\right\rangle $ with an energy $E_{F}$; the second is the transition
of the atom (molecule) from the state $\left| F\right\rangle $ to the final
state $\left| F^{\prime }\right\rangle =\left| N_{0}\right\rangle $ and
radiation of the photon with the momentum $\ \overrightarrow{k}^{\prime }=%
\overrightarrow{k}$.

Let $H_{A}$\ is the atom (molecule) Hamiltonian considering the weak
interaction of electrons and nucleus and the electromagnetic interaction of
an atom (molecule) with the external electric field $\overrightarrow{E}$. It
determines the system of eigenfunctions $\left| F\right\rangle $\ and
eigenvalues $E_{F}$ : 
\begin{equation}
H_{A}\left| F\right\rangle =E_{F}\left| F\right\rangle .
\end{equation}

The matrix element of the process determining the scattering amplitude in
forward direction in the electric dipole approximation can be written as 
\cite{8}: 
\begin{equation}
\frak{M}=\sum_{F}\left\{ \frac{\left\langle N_{0}\right| \overrightarrow{d}%
\overrightarrow{e}^{\ast }\left| F\right\rangle \left\langle F\right| 
\overrightarrow{d}\overrightarrow{e}\left| N_{0}\right\rangle }{%
E_{F}-E_{N_{0}}-\hbar \omega }+\frac{\left\langle N_{0}\right| 
\overrightarrow{d}\overrightarrow{e}\left| F\right\rangle \left\langle
F\right| \overrightarrow{d}\overrightarrow{e}^{\ast }\left|
N_{0}\right\rangle }{E_{F}-E_{N_{0}}+\hbar \omega }\right\} ,
\end{equation}
where $\overrightarrow{d}$\ is the electric dipole transition operator, $%
\omega $\ is the photon frequency.

Let us remind, that the atom (molecule) exited energy levels are
quasistationary: $E_{F}=E_{F}^{(0)}-\frac{i}{2}\Gamma _{F}$\ , $E_{F}^{(0)}$%
\ is the atom (molecule) level energy, \ $\Gamma _{F}$ is the level width.

The matrix element (8) can be written as: 
\begin{equation}
\frak{M}=\alpha _{ik}e_{i}^{\ast }e_{k},
\end{equation}
where the tensor of dynamical atom (molecule) polarizability $\alpha _{ik}$\
has the form

\begin{equation}
\alpha _{ik}=\sum_{F}\left\{ \frac{\left\langle N_{0}\right| d_{i}\left|
F\right\rangle \left\langle F\right| d_{k}\left| N_{0}\right\rangle }{%
E_{F}-E_{N_{0}}-\hbar \omega }+\frac{\left\langle N_{0}\right| d_{k}\left|
F\right\rangle \left\langle F\right| d_{i}\left| N_{0}\right\rangle }{%
E_{F}-E_{N_{0}}+\hbar \omega }\right\}
\end{equation}

The tensor $\alpha _{ik}$ can be expanded over the irreducible parts as 
\begin{equation}
\alpha _{ik}=\alpha _{0}\delta _{ik}+\alpha _{ik}^{s}+\alpha _{ik}^{a},
\end{equation}
where $\alpha _{0}=\frac{1}{3}\underset{i}{\sum \alpha _{ii}}$ is the
scalar, $\alpha _{ik}^{s}=\frac{1}{2}(\alpha _{ik}+\alpha _{ki})-$\ $\alpha
_{0}\delta _{ik}$\ is the symmetrical tensor of rank two, $\alpha _{ik}^{a}=%
\frac{1}{2}(\alpha _{ik}-\alpha _{ki})$ is the antisymmetrical tensor of
rank two. 
\begin{equation}
\alpha _{ik}^{a}=\frac{\omega }{\hbar }\sum_{F}\left\{ \frac{\left\langle
N_{0}\right| d_{i}\left| F\right\rangle \left\langle F\right| d_{k}\left|
N_{0}\right\rangle -\left\langle N_{0}\right| d_{k}\left| F\right\rangle
\left\langle F\right| d_{i}\left| N_{0}\right\rangle }{\omega
_{FN_{0}}^{2}-\omega ^{2}}\right\} \text{,}
\end{equation}
where $\omega _{FN_{0}}=\dfrac{E_{F}-E_{N_{0}}}{\hbar }$.

Let atoms (molecules) are nonpolarized. The antisymmetrical part of
polarizability (12) is equal to zero in the absence of T- and P- odd
interactions. Let us remind that for P-odd and T-even interactions the
antisymmetrical part of polarizability differs from zero when we consider
both the electric and magnetic dipole transitions only \cite{7}.

We can calculate the antisymmetrical part $\alpha _{ik}^{a}$\ of the tensor
of dynamical atom (molecule) polarizability $\alpha _{ik}$, and as a result,
to obtain the expression for $\beta _{E}^{T}$\ by the following way.
According to (4,6) the magnitude of the T-odd effect is determined by the
polarizability $\beta _{E}^{T}$\ or (that is actually the same, see (2,6))
by the amplitude $f_{\pm }(0)$\ of elastic coherent scattering of a photon
by an atom (molecule). When $\overrightarrow{n}_{E}\parallel \overrightarrow{%
n}_{\gamma }$\ the amplitude $f_{\pm }(0)$\ in dipole approximation can be
written as $f_{\pm }=\mp \dfrac{\omega ^{2}}{c^{2}}\beta _{E}^{T}$. As a
result, in order to obtain the amplitude $f_{\pm }$\ , the matrix element
(8,9) for photon polarization states $\overrightarrow{e}=\overrightarrow{e}%
_{\pm }$\ should be find.

\ The electric dipole transition operator$\overrightarrow{d}$ can be written
in the form: 
\begin{equation}
\overrightarrow{d}=d_{+}\overrightarrow{e}_{+}+d_{-}\overrightarrow{e}%
_{-}+d_{z}\overrightarrow{n}_{\gamma }\text{,}
\end{equation}
with $\overrightarrow{d}_{+}=-\dfrac{d_{x}-id_{y}}{\sqrt{2}}$, $%
\overrightarrow{d}_{-}=\dfrac{d_{x}+id_{y}}{\sqrt{2}}$.

Let $\overrightarrow{e}=\overrightarrow{e}_{+}$. Using (8,9) we can write
for polarizability $\beta _{E}^{T}$ the following expression: 
\begin{equation}
\beta _{E}^{T}=\frac{\omega }{\hbar }\sum_{F}\left\{ \frac{\left\langle
N_{0}\right| d_{-}\left| F\right\rangle \left\langle F\right| d_{+}\left|
N_{0}\right\rangle -\left\langle N_{0}\right| d_{+}\left| F\right\rangle
\left\langle F\right| d_{-}\left| N_{0}\right\rangle }{\omega
_{FN_{0}}^{2}-\omega ^{2}}\right\} .
\end{equation}

The more detailed expression for atom (molecule) wave functions is necessary
for the further calculations of the polarizability. The weak interactions
constant is very small. Therefore, we can use the perturbation theory. Let $%
\left| f,E\right\rangle $ is the wave function of an atom (molecule)
interacting with an electric field $\overrightarrow{E}$\ in the absence of
weak interactions $(V_{w}=0)$. Switch on weak interaction $(V_{w}\neq 0)$.
According to the perturbation theory\ \cite{8}\ the wave function $\left|
F\right\rangle $\ in this case can be written as: 
\begin{equation}
\left| F\right\rangle =\left| f,\overrightarrow{E}\right\rangle +\sum_{n}%
\frac{\left\langle n,\overrightarrow{E}\right| V_{w}\left| f,\overrightarrow{%
E}\right\rangle }{E_{f}-E_{n}}\left| n,\overrightarrow{E}\right\rangle
\end{equation}

From (14,15) for the polarizability $\beta _{E}^{T}$\ we obtain : 
\begin{equation}
\beta _{E}^{T}=\frac{\omega }{\hbar }\sum_{f}\frac{1}{\omega
_{fn_{0}}^{2}-\omega ^{2}}\sum_{l}
\end{equation}

\bigskip $\left\{ \frac{2\func{Re}\left[ \left\{ \left\langle n_{0}%
\overrightarrow{E}\right| d_{-}\left| f\overrightarrow{E}\right\rangle
\left\langle f\overrightarrow{E}\right| d_{+}\left| l\overrightarrow{E}%
\right\rangle -\left\langle n_{0}\overrightarrow{E}\right| d_{+}\left| f%
\overrightarrow{E}\right\rangle \left\langle f\overrightarrow{E}\right|
d_{-}\left| l\overrightarrow{E}\right\rangle \right\} \left\langle l%
\overrightarrow{E}\right| V_{w}\left| n_{0}\overrightarrow{E}\right\rangle %
\right] }{E_{n_{0}}-E_{l}}\right. +$

\bigskip

$\left. +\frac{2\func{Re}\left[ \left\langle n_{0}\overrightarrow{E}\right|
d_{-}\left| l\overrightarrow{E}\right\rangle \left\langle l\overrightarrow{E}%
\right| V_{w}\left| f\overrightarrow{E}\right\rangle \left\langle f%
\overrightarrow{E}\right| d_{+}\left| n_{0}\overrightarrow{E}\right\rangle
-\left\langle n_{0}\overrightarrow{E}\right| d_{+}\left| l\overrightarrow{E}%
\right\rangle \left\langle l\overrightarrow{E}\right| V_{w}\left| f%
\overrightarrow{E}\right\rangle \left\langle f\overrightarrow{E}\right|
d_{-}\left| n_{0}\overrightarrow{E}\right\rangle \right] }{E_{f}-E_{l}}%
\right\} $

It should be noted that the radial parts of the atom wave functions are real 
\cite{9}, therefore the matrix elements of the operators $d_{\pm }$ are real
too.

As a result, we can conclude that the P-odd T-even part of the interaction $%
V_{w}$\ does not give any contribution to $\beta _{E}^{T}$\ \ because the
P-odd T-even matrix elements for $V_{w}$\ are imaginary \cite{7}.\ At the
same time, the T- and P-odd matrix elements for $V_{w}$\ are\ real \cite{7},
therefore, the polarizability $\beta _{E}^{T}\neq 0$. It should be mentioned
that in the absence of electric field ($\overrightarrow{E}=0$) the
polarizability $\beta _{E}^{T}=0$\ \ and, therefore, the phenomenon of the
photon polarization plane rotation is absent.

Really, an electric field $\overrightarrow{E}$\ mixes atom levels with the
opposite parity. The atom levels have the fixed parity when $\overrightarrow{%
E}=0$. The operators\ $d_{\pm }$\ and $V_{w}$\ change the parity of the atom
states. As a result, the parity of a final state $\left| N_{0}^{\prime
}\right\rangle =$\ \ $d_{+}$\ $d_{-}$\ $V_{w}$\ $\left| N_{0}\right\rangle $%
\ appears to be opposite to the parity of initial state $\left|
N_{0}\right\rangle $. But the initial and final states \ in the expression
for $\beta _{E}^{T}$\ are the same. Therefore $\beta _{E}^{T}$\ can not
differ from zero\ when $\overrightarrow{E}=0$.

Let us estimate the magnitude of the T-odd photon plane rotation effect now.
According to the analysis \cite{4,5,6}, based on the calculations \ of the
value of \ T and P noninvariant interactions given by \cite{7} the ratio $%
\dfrac{\left\langle V_{w}^{T}\right\rangle }{\left\langle
V_{w}^{P}\right\rangle }\leq 10^{-3}\div 10^{-4}$, where\ $\left\langle
V_{w}^{T}\right\rangle $\ is T and P-odd matrix element, $\left\langle
V_{w}^{P}\right\rangle $\ is P-odd T-even matrix element.

The P-odd T-even polarizability $\beta _{s}^{P}$\ is proportional to the
electric dipole matrix element, the magnetic dipole matrix element and $%
\left\langle V_{w}^{P}\right\rangle $: $\beta _{s}^{P}\sim $\ $\left\langle
d\right\rangle \left\langle \mu \right\rangle \left\langle
V_{w}^{P}\right\rangle $\ \cite{7}. At the same time $\beta _{E}^{T}\sim $\ $%
\left\langle d(\overrightarrow{E})\right\rangle \left\langle d(%
\overrightarrow{E})\right\rangle \left\langle V_{w}^{T}\right\rangle $. As a
result 
\begin{equation}
\frac{\beta _{E}^{T}}{\beta _{s}^{P}}\sim \frac{\left\langle d(%
\overrightarrow{E})\right\rangle \left\langle d(\overrightarrow{E}%
)\right\rangle \left\langle V_{w}^{T}\right\rangle }{\left\langle
d\right\rangle \left\langle \mu \right\rangle \left\langle
V_{w}^{P}\right\rangle }.
\end{equation}

Let us study the T-odd phenomena of the photon polarization plane rotation
in an electric field $\overrightarrow{E}$\ for the transition $%
n_{0}\rightarrow f$ \ between the levels $n_{0}$ and $f$\ which have the
same parity when the field $\overrightarrow{E}=0$.\textbf{\ }In this case%
\textbf{{\Large \ } }the matrix element\ $\left\langle n_{0},\overrightarrow{%
E}\right| d_{\pm }\left| f,\overrightarrow{E}\right\rangle $\ does not equal
to zero for $\overrightarrow{E}\neq 0$ only. Let the interaction energy of
an atom with an electric field $V_{E}=-$ $\overrightarrow{d}\overrightarrow{E%
}$\ is much smaller then the spacing $\Delta $\ of the energy levels, which
are mixed by the field $\overrightarrow{E}$. It allows us to use the
perturbation theory for the wave functions $\left| f,\overrightarrow{E}%
\right\rangle $: 
\begin{equation}
\left| f,\overrightarrow{E}\right\rangle =\left| f\right\rangle +\sum_{m}%
\frac{\left\langle m\right| -d_{z}E\left| f\right\rangle }{E_{f}-E_{m}}%
\left| m\right\rangle ,
\end{equation}

where $z$ axis is parallel to\ $\overrightarrow{E}$.

As a result, the matrix element $\left\langle n_{0},\overrightarrow{E}%
\right| d_{\pm }\left| f,\overrightarrow{E}\right\rangle $\ \ can be written
as: 
\begin{eqnarray}
\left\langle n_{0},\overrightarrow{E}\right| d_{\pm }\left| f,%
\overrightarrow{E}\right\rangle &=&  \notag \\
&=&-\left\{ \sum_{m}\frac{\left\langle n_{0}\right| d_{\pm }\left|
m\right\rangle \left\langle m\right| d_{z}\left| f\right\rangle }{E_{f}-E_{m}%
}+\right. \\
&&+\left. \sum_{p}\frac{\left\langle n_{0}\right| d_{z}\left| p\right\rangle
\left\langle p\right| d_{\pm }\left| f\right\rangle }{E_{n_{0}}-E_{p}}%
\right\} E.  \notag
\end{eqnarray}

One can see that the matrix element $\left\langle d(\overrightarrow{E}%
)\right\rangle \sim \ \dfrac{\left\langle d\right\rangle E}{\Delta }$\ $%
\left\langle d\right\rangle $\ in this case.

The others matrix elements in (16) we can calculate for $\overrightarrow{E}%
=0 $.

As a result, we can obtain the estimation 
\begin{equation}
\beta _{E}^{T}\sim \left\langle d\right\rangle \left\langle d\right\rangle 
\frac{\left\langle dE\right\rangle }{\Delta }\left\langle
V_{w}^{T}\right\rangle \ \ .
\end{equation}

The ratio (17) can be written as 
\begin{equation}
\frac{\beta _{E}^{T}}{\beta _{s}^{P}}\sim \frac{\left\langle d\right\rangle
\left\langle d\right\rangle \frac{\left\langle dE\right\rangle }{\Delta }%
\left\langle V_{w}^{T}\right\rangle }{\left\langle d\right\rangle
\left\langle \mu \right\rangle \left\langle V_{w}^{P}\right\rangle }.
\end{equation}

The matrix element $\left\langle \mu \right\rangle \sim \alpha \left\langle
d\right\rangle $\ \ \cite{8,9}, where $\alpha =\frac{1}{137}$ is the fine
structure constant. Finally we have: 
\begin{equation}
\frac{\beta _{E}^{T}}{\beta _{s}^{P}}\sim \alpha ^{-1}\frac{\left\langle
dE\right\rangle }{\Delta }\frac{\left\langle V_{w}^{T}\right\rangle }{%
\left\langle V_{w}^{P}\right\rangle }
\end{equation}

For the case $\dfrac{\left\langle dE\right\rangle }{\Delta }\sim 1$ the
ratio (22) gives 
\begin{equation}
\frac{\beta _{E}^{T}}{\beta _{s}^{P}}\sim \alpha ^{-1}\frac{\left\langle
V_{w}^{T}\right\rangle }{\left\langle V_{w}^{P}\right\rangle }\lesssim
10^{-1}\div 10^{-2}
\end{equation}

This is possible, for example, for exited states of atoms and for two-atom
molecules (TlF, BiS, HgF). As one can see, the ratio $\dfrac{\beta _{E}^{T}}{%
\beta _{s}^{P}}$ is two orders larger as compared with the simple estimation 
$\dfrac{\left\langle V_{w}^{T}\right\rangle }{\left\langle
V_{w}^{P}\right\rangle }\leq 10^{-3}\div 10^{-4}$ due to the fact that $%
\beta _{E}^{T}$\ is determined by the electric dipole transitions only,
while $\beta _{s}^{P}$\ is determined both by the electric and magnetic
dipole transitions. As a result, the experimental observation of the T-odd
photon polarization plane rotation becomes real. The effect can be increased
by using a resonator or a volume diffraction grating \cite{6}.

The absorption of the photon in the matter does not allow to do the length $%
L $\ significantly greater then the absorption length $L_{a}$. This
difficulty can be removed when the photon passes in the medium with the
inverse population of the atom states (the conditions of laser generation or
amplification are fulfilled for the transitions of our interest). The
coherent wave is not absorbed in this case, it is amplified. As a result the
length $L$\ can be done significantly greater{\Large \ }then $L_{a}$. We can
use the high quality resonators and diffraction gratings for the effective
enhancement of the length $L$. Let us note that due to T-odd interactions
the amplification depends on the photon circular polarization sign. So we
can study two effects in laser medium: T-odd photon plane rotation in the
electric field $\overrightarrow{E}$\ and circular dichroism - different
amplification for the left and right photons. The atoms Cs, Tl, Pb, Dy are
very convenient\ for this as well as for P-odd T-even phenomena studying 
\cite{4,5,6}. It is important, that for such atoms laser generation has been
done.

5. Thus we have shown that the T-odd and P-odd phenomena of photons
polarization plane rotation and circular dichroism are not very small in
comparison with P-odd T-even effects. As a result, we can hope to measure
them experimentally. It is convenient to use atoms Cs, Tl, Pb, Dy and
two-atom molecules containing heavy atom TlF, BiS, HgF, DyF for the
investigation of P,T-odd phenomena as well as for the investigation of P-odd
T-even\ ones \cite{7}.

Let us note, that the new T-odd and P-odd phenomena of photon polarization
plane rotation (circular dichroism) in an electric field has general
meaning. So, the super strong electric fields are formed in the collisions
of the relativistic nuclei. The $\gamma $-quanta originating in such
collisions propagate in the nuclear matter which interacts with the electric
field. As a result the radiated\ $\gamma $-quanta have the circular
polarization. The gravitational field stipulates the appearance of the
gravitational analog of the Stark mixing of atom (molecule) levels. It means
that the gravitational analog of the T-odd polarizability $\beta _{g}^{T}$\
exists.\ \ As a result, the T-odd P-odd phenomena of photon plane rotation
(circular dichroism) appears, for example, for an electromagnetic wave
moving in a matter interacting with the strong gravitational field of a star
(term in the scattering amplitude proportional to\ $\beta _{g}^{T}(%
\overrightarrow{g}[\overrightarrow{e}^{\ast }\times \overrightarrow{e}])$\ ,
\ $\overrightarrow{g}$\ is the free fall acceleration). Due to quantum
electrodynamic effects of electron-positron pair creation in strong electric
and magnetic fields  the dielectric permittivity tensor $\varepsilon _{ik}$
of vacuum depends on $\overrightarrow{E\text{ }}$ and $\overrightarrow{H}$ 
\cite{8}. The theory of \ $\varepsilon _{ik}$ \cite{8} does not take into
account the weak interaction of electron and positron with each other.
Considering the weak interactions of electron and positron with each other
in the process of pair creation in an electric field one can obtain, that
the permittivity tensor of vacuum in strong electric (gravitational) field
contains the term \ $\beta _{vac\overrightarrow{E}}^{T}(\overrightarrow{n_{E}%
}[\overrightarrow{e}^{\ast }\times \overrightarrow{e}])$\ , $\varepsilon
_{ik}^{vac}\sim i\beta _{vac\overrightarrow{E}}^{T}\varepsilon _{ikl}n_{lE}$
($\beta _{vac\overrightarrow{g}}^{T}(\overrightarrow{g}[\overrightarrow{e}%
^{\ast }\times \overrightarrow{e}])$, $\varepsilon _{ik}^{vac}\sim i\beta
_{vac\overrightarrow{g}}^{T}\varepsilon _{ikl}\overrightarrow{g}$), and as a
result, the polarization plane rotation (circular dichroism) phenomena exist
for photon moving in an electric (gravitational) field in vacuum. And visa
versa\ $\gamma $-quanta appeared \ while single-photon electron-positron
annihilation in an electric (gravitational) field will have the admixture of
circular polarization, caused by T-odd P-odd weak interactions.

\end{document}